\documentclass[showpacs,twocolumn,aps,floatfix,superscriptaddress,endnotes,nofootinbib]{revtex4-1}
\usepackage{amsmath,amssymb,eucal,graphicx,bm,verbatim}
\usepackage[dvips]{color}

\begin{document}
\title{Universal energy fluctuations in thermally isolated driven systems}

\author{Guy Bunin*}
\affiliation{Department of Physics, Technion, Haifa 32000, Israel}
\author{Luca D'Alessio*}
\affiliation{Department of Physics, Boston University, Boston, MA 02215, USA}
\author{Yariv Kafri}
\affiliation{Department of Physics, Technion, Haifa 32000, Israel}
\author{Anatoli Polkovnikov}
\affiliation{Department of Physics, Boston University, Boston, MA 02215, USA}
\footnotetext{These authors contributed equally to this project.\\Email Addresses:\\buning@tx.technion.ac.il\\dalessio@bu.edu\\kafri@physics.technion.ac.il\\asp@bu.edu}

\begin{abstract}
When an isolated system is brought in contact with a heat bath its final energy
is random and follows the Gibbs distribution -- a cornerstone of statistical physics.
The system's energy can also be changed by performing non-adiabatic work using a cyclic process.
Almost nothing is known about the resulting energy distribution in this setup, which
is especially relevant to recent experimental progress in cold atoms, ions traps, superconducting
qubits and other systems.
Here we show that when the non-adiabatic process comprises of many repeated cyclic
processes the resulting energy distribution is universal and different from the Gibbs ensemble.
We predict the existence of two qualitatively different regimes with a continuous
second order like transition between them.
We illustrate our approach performing explicit calculations for both interacting and non-interacting systems.
\end{abstract}
\maketitle

Understanding equilibrium and non-equilibrium properties of thermally \textit{isolated} systems has become a forefront of research due to experimental developments over the past decade, particularly in cold atom systems \cite{bloch_rmp}, trapped ions \cite{blatt_08}, and nuclear spins~\cite{petta_08} and superconducting qubits~\cite{majer_07}. In these systems the coupling to external dissipative degrees of freedom is strongly suppressed and irrelevant on accessible time scales. These systems provide a new and very clean playground where one can investigate fundamental questions in statistical and quantum physics. Moreover, they point to new practical applications, in particular in the context of quantum information.
The experimental studies inspired intensive theoretical research on a variety of topics. These include equilibration in isolated systems initially driven out of equilibrium by a sudden change in a coupling constant (a quench); defect (or energy) generation during slow nearly adiabatic processes in gapless phases or near singularities, such as quantum phase transitions (for a review see Refs.~\cite{dziarmaga_review, ap_rmp}); non-equilibrium quantum phase transitions in the presence of $1/f$ noise~\cite{ehud_1/f}; and many more.

\begin{figure}
\includegraphics[width=0.9\columnwidth]{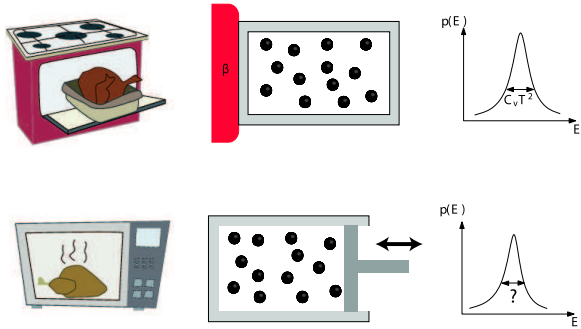}
\caption{Schematic comparison between the usual thermal heating (traditional oven, top) and an energy increase due to not-adiabatic work (microwave oven, bottom). On the right of each case we present a schematic picture of the resulting energy distribution in each case.}
\label{ovens2}
\end{figure}

In this work, we consider the energy distribution of a thermally {\it isolated} system following a non-adiabatic process. Consider two setups where the energy of an isolated system is changed. In the first, the system is brought into contact with a heat bath until equilibration and then disconnected from it - similar to an oven. In the second the energy of the system is increased due a non-adiabatic change of some external parameter(s) - much like a microwave. The two setups are illustrated schematically in Fig.~\ref{ovens2}. Our interest is in the energy distribution of each of the systems at the end of the process. For the first setup the result is well-known and corresponds to the classic heating mechanism which can be found in any book on thermodynamics (see e.g. Ref.~\cite{reif}). If, as usual, the bath is large compared to the system then the energy distribution of the system becomes Gaussian, with a canonical width uniquely determined by the fluctuation-dissipation relation: $\delta E^2=T^2 C_v$, where $T$ is the temperature and $C_v$ is the specific heat. This relation is valid for both quantum and classical systems and is independent of the details of the interactions between the system and the bath. Now, consider a second setup where the energy of the system is changed due to a non-adiabatic variation of an external parameter (say, the electro-magnetic field in the case of a microwave or the motion of the piston in Fig.~\ref{ovens2}). This type of heating also inevitably leads to an uncertainty in the final energy of the system. While it is known that the energy changes in the system obey, even beyond linear response, the recently discovered fluctuation theorems (see for example, ~\cite{jarzynski_97, crooks_98, hanggi_10}) very little is known about the resulting energy distribution in this case~\cite{NeriYariv}. In a large macroscopic system the energy distribution is  expected to be very narrow and the relative energy fluctuations negligible. But in small or mesoscopic systems, which are of primary experimental interest (see e.g. Ref.~\cite{bloch_rmp}), the fluctuations can be large and important. Fundamental questions are unanswered: Which features of the energy distribution are universal and which features depend on details of the system and driving protocol? To what extent can the width of the distribution be controlled? For example, can one dynamically increase the energy of an isolated system without increasing the uncertainty in the final energy? Can the fluctuation-dissipation relations, which determine the energy width in the oven-like setup, be extended to the microwave-like setup?

In this paper we begin to address such questions. Specifically, we study a thermally isolated system undergoing a repeated cyclic process, whereby some external parameter $\lambda$ in the Hamiltonian is changed in time and returns to its original value at the end of each cycle, see Fig.~\ref{ovens2}.
We show that under generic assumptions of (i) small work per cycle and (ii) absence of correlations between cycles (see detailed discussion in Methods)  the variance of the energy distribution $\sigma^2(E)$ at energy $E$ assumes a particularly simple form to leading order in $1/N$, where $N$ is the number of degrees of freedom in the system. It depends only on the microcanonical temperature $\beta(E)=\partial_E \ln\Omega(E)$ where $\Omega(E)$ is the density of states, and on the average energy change in a cycle at energy $E$, $A(E)$:
\begin{equation}
\label{widthbeta}
\sigma^{2}\left(E\right)=\sigma^{2}_0\frac{A^{2}\left(E\right)}{A^{2}\left(E_0\right)}
+ 2 A^{2}\left(E\right) \int_{E_{0}}^{E}\frac{dE^{\prime}}{A^{2}\left(E^{\prime}\right)\beta\left(E^{\prime}\right)}\;.
\end{equation}
Here $E_{0}$ is the initial energy of the system and $\sigma^{2}_0$ is the initial variance.
This equation is the main result of the paper. 

The result~(\ref{widthbeta}) follows from integrating a Fokker-Planck equation which describs the time evolution of the energy distribution $P(E,t)$ (see Methods and Supplementary Material for details):
\begin{equation}
\partial_{t}P=-\partial_{E}(A(E)P)+\frac{1}{2}\partial_{EE}(B(E)P).\label{FP_main}
\end{equation}
The change of the energy distribution in one cycle of the protocol is obtained by integrating this equation over the duration of the protocol, set for simplicity to be unity. Within this choice $A(E),B(E)$ represent the average work per cycle and its variance respectively: $A=\langle w\rangle $ and $B=\langle w^2 \rangle_c$. Here the angular brackets denote averaging over realizations of the cycle starting from a fixed initial energy.

In general $A(E)$ and $B(E)$ are protocol dependent functions and are {\em a priori} independent from each other. However, since the system is thermally isolated its time evolution is governed by Hamilton's equations of motion in the classical case and the Schrodinger equation in the quantum case. This puts strong constraints on the relation between $A(E)$ and $B(E)$ similar to the Einstein fluctuation-dissipation relations between drift coefficient (mobility) and diffusion in open systems~\cite{ashcroft-mermin}. In particular, we find
\begin{equation}
\beta B=2A-\partial_{E}B=2A+{\cal O}(N^{-1}).
\label{real_AB}
\end{equation}
For interacting systems with many degrees of freedom the second term on the RHS of this equation is a $1/N$ correction which can be neglected. This term can be important though in mesoscopic or integrable systems. Eq.~(\ref{real_AB}) was previously suggested for classical systems in Ref.~\cite{jarzynski_92}.  The fluctuation-dissipation relation~(\ref{real_AB}) is derived within a small work assumption (explicitly $\beta(E)^2 \langle w^3 \rangle_c \ll A(E)$, where $\langle w^3 \rangle_c$ is the third cumulant of the work, and $A(E)\ll TC_v$, see Methods and Supplementary Material). As we will show below Eq.~(\ref{real_AB}) holds for a very wide class of classical and quantum systems starting from noninteracting particles in a time dependent cavity to fully interacting spin systems. The main result of the paper Eq.~(\ref{widthbeta}) is a direct consequence of this relation (see Methods).

Several interesting consequences follow from Eq.~\eqref{widthbeta}: (i) When $A(E)$ is constant the energy width depends only on $\beta(E)$, and not on the amplitude of the drive or other details of the driving protocol. (ii) When $A(E)$ is not constant, depending on the functional form of $A(E)$ and $\beta(E)$, the variance of the distribution can be larger and surprisingly, even smaller than the width of the equilibrium Gibbs distribution at the same mean energy. In fact, $\sigma^{2}(E)/\sigma^{2}_{eq}(E)$ can be made arbitrarily small by a proper choice of $A(E)$. (iii) When $A$ is a function of the energy density $u=E/N$ (with a possible extensive energy independent prefactor like the total number of particles), we have $\sigma^{2}(E)\sim {\cal O} (N)$, scaling as in equilibrium. For a single quench this result was noticed e.g. in Ref.~\cite{silva_08}. Here we show that it remains valid after many quenches. (iv) The dependence of $\sigma^2$ on $E$ displays two qualitatively distinct behaviors with increasing $E$, depending on whether the integral in Eq.~(\ref{widthbeta}) diverges or converges as $E \to \infty$.

To illustrate the distinct behaviors associated with point (iv) above we consider
the generic case where $\beta\propto E^{-\alpha}$,
which is the case for phonons, superfluids or other systems with Goldstone bosons, Fermi liquids, ideal gases and others. Moreover we measure time in units of the number of cycles carried out (in what follows we will use time and number of cycles interchangeably) and assume a simple power law behavior for $A(E)$:
\begin{equation}
\label{s}
\partial_t E=A(E)=c E^s \;.
\end{equation}
As will become clear below the two regimes exist even in cases when $A(E)$ and $\beta(E)$ are not power laws. The values of $\alpha$ are constrained by simple thermodynamic arguments to $0 <\alpha\le1$: the lower bound is required by positivity of the specific heat and the upper bound assures that the entropy ($S(E)\propto E^{1-\alpha}$) is an increasing unbounded function of the energy. To prevent the system's energy from diverging in a finite time we require $s\le1$ (as follows from integrating Eq.~\eqref{s}).

For simplicity we also assume $\sigma_{0}(E_{0})= 0$ and compare the width to the equilibrium canonical width $\sigma_{eq}^{2}=-\partial_\beta E\sim E^{1+\alpha}/\alpha$. In this case the system displays a {\it transition} between two behaviors as the functional form of $A(E)$ is changed. This transition is continuous and is characterized by a diverging time-scale needed to reach the asymptotic regime. Specifically, depending on the sign of $\eta = 2s-1-\alpha$, Eq.~(\ref{widthbeta}) implies: (i) When $\eta < 0$ the width is \textit{Gibbs-like} with $\sigma^2/\sigma_{\rm eq}^2\to 2\alpha/|\eta|$, i.e. the ratio $\sigma^{2}/\sigma_{eq}^{2}$ asymptotically approaches a constant value that can be either larger or smaller than one. Note that smaller widths correspond to protocols with large and negative $s$, i.e. to protocols where $A(E)$ is a strongly decreasing function of energy. (ii) A second \textit{run-away} regime occurs when $\eta>0$. Here the width increases with a higher power of energy than the canonical width: $\sigma^{2}/\sigma_{eq}^{2}\sim E^{\eta}$. The resulting distribution is significantly wider than the canonical one. Given the constraint on the value of $s$, this regime can only be reached if $\alpha<1$, in particular this regime is unreachable for a classical ideal gas.
The transition between the \textit{Gibbs-like} and \textit{run-away} regimes occurs when  $\eta=0$ which implies $\sigma^{2}/\sigma_{eq}^{2}\sim 2 \alpha \ln \left( \frac{E}{E_0}\right)$. Close to this transition when $|\eta|\ll 1$, there is a divergent time scale (or number of cycles) required to reach the asymptotic regime. This time scale can be obtained by combining Eq.~\eqref{widthbeta} and Eq.~\eqref{s}, see Table~\ref{summary}. The crossover from the {\it Gibbs-like} to the {\it run-away} regime is qualitatively similar to a continuous phase transition, with the diverging time scale being analogous to a divergent relaxation time (critical slowing down) in the equilibrium case. We summarize our results, close to the transition, for the above choices of $\beta(E)$ and $A(E)$ in Table~\ref{summary}.

The qualitative difference between the two regimes can also be understood in terms of the entropy of the distribution: ${\cal S}=-\sum_n \rho_n\ln \rho_n$, where $\rho_n$ are the microscopic probabilities to occupy different energy levels. Converting this sum into an integral over energies and expanding the resulting expression up to $1/N$ corrections it is straightforward to check~\cite{santos_11} that:
\begin{displaymath}
{\cal S}(E)-{\cal S}_{\rm eq} (E)=\ln\left(\sigma(E)\over \sigma_{\rm eq}(E)\right)+{1\over 2}\left(1-{\sigma^2(E)\over \sigma_{\rm eq}^2(E)}\right),
\end{displaymath}
where ${\cal S}_{\rm eq}(E)=\ln(\sqrt{2\pi}\sigma_{\rm eq}(E)\Omega(E))$ is the equilibrium canonical entropy. It is easy to see that the correction to the equilibrium entropy is always negative except when the width of the energy distribution coincides with the canonical width. In the {\it Gibbs-like} regime this correction is a constant, while in the {\it run-away} regime it has an explicit energy dependence.

Note that under the assumptions used to derive our main result, the integral in Eq.~\eqref{widthbeta} can be rewritten in terms of $\widehat{A}(S)$ the average entropy change per unit of time (per-cycle), where $S=\ln \Omega(E)$. Using $A(E)=\widehat{A}(S)\partial_{S} E$ the nature of the transition between the two regimes assumes an interesting physical interpretation. Specifically, the integral in Eq.~(\ref{widthbeta}) becomes $\int dS/\widehat{A}^2(S)$ showing that the {\em Gibbs-like} and ({\em run-away}) regimes correspond to the entropy growing slower (faster) than time squared. This conclusion does not rely on any assumptions about the specific functional form of $\beta(E)$ and $A(E)$.

\begin{table}
\begin{center}
\begin{tabular}{|c|c|c|c|}
\hline
Regime & Condition & Time Scale & width\tabularnewline
\hline
\textit{Gibbs-like} & $\eta<0$ & $\frac{E_0^{1-s}}{c}\frac{1}{(1-s)}\exp[\frac{1-s}{|\eta|}]$
& $\frac{\sigma^{2}}{\sigma_{eq}^{2}}\sim\frac{2\alpha}{|\eta|}$
\tabularnewline
\hline
\textit{run-away} & $\eta>0$ & $\frac{E_0^{1-s}}{c}\frac{1}{(1-s)}\exp[\frac{1-s}{\eta}]$
& $\frac{\sigma^{2}}{\sigma_{eq}^{2}}\sim\frac{2\alpha}{\eta}\left(\frac{E}{E_{0}}\right)^{\eta\phantom{X}}$\tabularnewline
\hline
\textit{critical} & $\eta=$0 & - & $\frac{\sigma^{2}}{\sigma_{eq}^{2}}\sim2\alpha\log\left(\frac{E}{E_{0}}\right)$\tabularnewline
\hline
\end{tabular}
\hfill{}
\caption{A summary of the results
 for $A(E)=c E^s$, $\beta\sim E^{-\alpha}$ with $0<\alpha\le1$, $s\le1$
and $\eta=2s-1-\alpha$. The width specifies the asymptotic value in units of the equilibrium width at the same energy. The time scale specifies the characteristic ``relaxation'' time needed to reach the asymptotic regime.}
\label{summary}
\end{center}
\end{table}

\section{Examples}
First, we consider a system of non-interacting and weakly interacting particles in a deforming cavity. Then we analyze a single particle in a harmonic potential, which is a part of a larger system, and subject to a time-dependent external force. Two additional examples of a classical one-dimensional XY-model and a quantum one-dimensional transverse field Ising model will be discussed in the Supplementary Material.

\begin{figure}
\includegraphics[width=0.9\columnwidth]{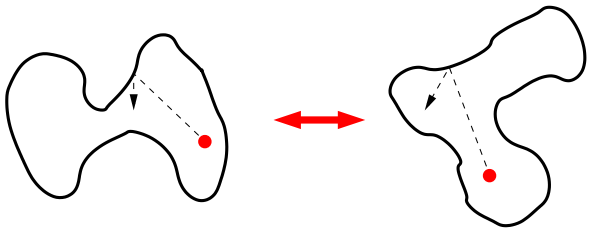}
\caption{An illustration of a single particle bouncing in a deforming cavity of constant volume.
The driving protocol consists in repeatedly deforming the cavity between the two shapes shown.}
\label{jar2}
\end{figure}

{\it Single Particle in a deforming cavity}: Let us first consider a very simple system - a single particle bouncing elastically in a cavity. When the cavity is stationary the energy of the particle is conserved. If the cavity is chaotic there are no other conserved quantities so that in the long-time limit the particle relaxes to a uniform position distribution and an isotropic momentum distribution. We consider a process where the system is repeatedly driven by deforming the cavity. At the end of each cycle the cavity comes back to its original shape and the system is allowed to relax in the sense described above (see Fig.\ref{jar2}). In this setup the number of degrees of freedom $N$ is given by $N=2d$ where $d$ is the dimensionality of the system. In a single collision with the moving wall the particle's kinetic energy can either increase or decrease. However, it will always increase on average and eventually the particles velocity will become much greater than the velocity of the wall. Then the work per cycle automatically becomes small and the conditions for the validity of the Fokker-Planck equation are satisfied.

If the cavity is deformed while keeping its volume fixed then a very simple behavior emerges.
In this case it has been shown \cite{box1-th,box2-th,box-exp1,box-exp2} that the particles velocity distribution becomes exponential irrespective of the container's shape and the deformation protocol, $f(v,\tau)d{\bf v}\sim e^{-v/\tau}d{\bf v}$, where $\tau\sim \langle u^2\rangle t$ is proportional to time (number of cycles) and to the second moment of the velocity of the wall. Moreover if the cavity is sufficiently chaotic successive collisions are uncorrelated and the formalism holds even if the waiting time between cycles approaches zero (see \cite{box2-th} for details). In this case the fundamental equation for the energy distribution, $P(E)$, assumes the Fokker-Planck form of Eq.~\eqref{FP_main}
with $A(E)=c\,E^{1/2}$ and $B(E)=c\, 4 E^{3/2}/(d+1)$ where $c$ contains information about the mass of the particle, the area or volume of the container and the velocity of the moving walls (see Ref.~\cite{box2-th} for details). Using $\beta(E)=(d-2)/(2E)$ it is easy to verify that Eq.~\eqref{real_AB} is exactly satisfied with the $1/N$ correction included. Therefore, in the large $d$-limit, Eq.~\eqref{widthbeta} holds, and we find $\sigma^2(E)/\sigma^2_{eq}(E)\to 2$, consistent with the exponents $\alpha=1$ and $s=1/2$. With the $1/N$ correction included we find:
\begin{equation}
{\sigma^2(E)\over \sigma^2_{eq}(E)}=\frac{2+3/d}{1+1/d}.
\end{equation}
This result also follows from the exactly known single particle distribution~\cite{box1-th}. This result was also derived for a single light impurity moving in a background of heavy atoms (Lorenz gas). In this case a full microscopic description based on a Lorentz-Boltzmann equation is possible~\cite{LP}.

{\it Weakly-interacting particles}: Extending the above example, consider $n$ weakly interacting particles in a deforming cavity so that $N=2nd$. We assume that during each cycle the particles can be treated as non-interacting, while between cycles the system rethermalizes at a fixed total energy, so that the velocity distribution of the particles becomes Maxwell-Boltzmann rather than exponential. A calculation similar to that of ~\cite{box1-th, box2-th} shows that $A(E)$ and $B(E)$ have the same functional form as in the non-interacting case but with different prefactors. Moreover both $A(E)$ and $B(E)$ become extensive and the constraint, Eq.~\eqref{real_AB}, is satisfied with the $1/N$ correction becoming negligible.  Then Eq.~(\ref{widthbeta})  gives the asymptotic result $\sigma^2(E)/\sigma^2_{eq}(E)\to 2$, consistent with the exponents $\alpha=1$ and $s=1/2$. This result is identical to the single particle result in the large $N$ limit, despite the very different single particle velocity distributions.

In either noninteracting or interacting setups the two functions, $A(E),B(E)$, can be experimentally obtained by measuring the first two cumulants of the work distribution in one cycle: $\langle w \rangle=A(E)$ and $\langle w^2 \rangle_c=B(E)$. Alternatively one can measure the average energy and its variance versus time and determine $A(E)$ and $B(E)$ as the slopes of these two functions respectively.

{\it Single particle in a time-dependent potential}: Next we consider a classical particle in a harmonic trap, which is part of a larger system, e.g. a set of $N$ identical particles, whose details define $\Omega(E)$ and hence $\beta(E)$. We assume that the coupling to the rest of the system is weak and unimportant within the duration of a cycle, much like in the weakly-interacting particle gas example above. In contrast with the previous example, this setup illustrates driving a system with a local perturbation (which can, however, be applied independently to many different particles). The particle's energy $\varepsilon$ between cycles is given by
\[
\varepsilon=\frac{1}{2}kx^{2}+\frac{1}{2}mv^{2}.
\]
For simplicity we work in one dimension. For large $N$ the probability distribution for $\left(x,v\right)$ before the drive is $\rho(x,v)\propto\exp\left(-\beta(E) \varepsilon\right)$. We consider a driving process which consists of an impulse of magnitude $F\left(x\right)\Delta t$ with $\Delta t$ short enough so that the particle's position does not change appreciably during the drive. This assumption also guaranties that the coupling to the rest of the system is unimportant during a cycle. Under this impulse the velocity changes according to $v\rightarrow v+F\left(x\right)\Delta t$.
$A,B$ are readily calculated and read
\begin{align*}
A & =\left\langle \left(F\left(x\right)\right)^{2}\right\rangle \Delta t^{2}, \\
B & =\frac{2}{\beta}\left\langle \left(F\left(x\right)\right)^{2}\right\rangle \Delta t^{2},\;
\end{align*}
verifying fluctuation-dissipation relations (\ref{real_AB}). Taking $\beta\left(E\right)\propto E^{-\alpha}$ and $F\left(x\right)\propto x^{r}$
we find
\[
A\propto\left\langle x^{2r}\right\rangle \propto \epsilon^{r}\propto E^{\alpha r}.
\]
Using our previous convention, $A\propto E^{s}$, we see that $s=\alpha r$. Different values
of $\eta=2\alpha r-1-\alpha$ can be obtained using different impulse forces and systems. For example, for a system with
$\alpha=1/2$, such as a Fermi liquid or a one-dimensional harmonic system, with $r=1$ we obtain $\eta=-1/2$ leading to the {\it Gibbs-like} regime with $\sigma^2/\sigma_{eq}^2=2$. When $r=3/2$ we are at the critical regime $\eta=0$. Finally for $r=2$ we obtain $\eta=1/2$ leading to the {\em run-away} regime where $\sigma^2/\sigma_{eq}^2  \sim E^{1/2}$.

\section{Conclusions}

The main result of our paper is based on a fluctuation-dissipation relation connecting drift and diffusion of the energy in a driven system (Eq.~(\ref{real_AB})). This relation is very closely connected to the recently discovered fluctuation theorems. In fact, in the supplementary material we give a rigorous derivation of Eq.~(\ref{real_AB}) using the quantum version of fluctuation theorems, which we extend to our setup of repeated cyclic processes. In the original formulation the Jarzynski relation states that if a system starts from a {\em Gibbs distribution} the change in energy of the system, $w$, for cyclic processes obeys the equality $\langle e^{-\beta w} \rangle=1$. Here the angular brackets denote an average over both different realizations of the process and different initial conditions. As emphasized in Ref. \cite{NeriYariv} these relations holds very little information about the first few moments of the distribution of $w$, unless $w$ is small. Only when it is small, a cumulant expansion of the Jarzynski relation up to the second order in $w$ recovers the fluctuation-dissipation relation (\ref{real_AB}) without the $1/N$ correction (see Supplementary Material for details). When the energy changes are large its moments are governed by details of the physical process which have to be accounted for (see \cite{NeriYariv}). In this paper we (i) importantly, overcome the restriction of small energy changes by considering a large change which is a results of many small changes (leading to Eq.~(\ref{widthbeta})) and (ii) show that relation (\ref{real_AB}) is independent of the exact form of the initial distribution.

We believe that some of the assumptions of the work can be further relaxed. In particular, it can be shown that relation~(\ref{real_AB}) is valid for generic (non-cyclic) quasi-static process where the system is approximately in a steady state at each moment of time. In this case by $A$ is related to the non-adiabatic part of the work $w$. Likewise it is plausible that the assumptions of complete relaxation to a steady state between cycles are not necessary, at least for ergodic systems. Physically these assumptions amount to a loss of correlations between different cycles which is inevitable in ergodic systems. The assumption of unitary dynamics can be also relaxed. For example, we can allow measurements of the energy during the protocol, which in quantum systems project the system to one of the energy eigenstates. Such measurements do not invalidate the derivation given in Appendix \ref{extramaterial2}. These points will be addressed in a future study.

The predictions of this paper can be experimentally tested in cold atom systems or in driven nuclear spins. For example, a trapping potential can be modulated to perform non-adiabatic work on the system. The resulting energy distribution can then be probed using time of flight experiments. Likewise one can measure energy fluctuations due to a $1/f$ noise in systems of neutral atoms near atom chips or in trapped ions in a setup similar to that discussed in Ref.~\cite{ehud_1/f}. If there is no external cooling or dissipation in the system after its initial preparation (as is often the case) then the energy fluctuations induced by the noise should also agree with our predictions.

 Finally, we comment that energy fluctuations can be measured indirectly through averages or fluctuations of other observables, like the magnetization or correlation functions. In particular, it is easy to check that the expectation value of every observable $O$ and its variance, up to $1/N^2$ corrections, are given by
 \begin{displaymath}
 \langle O \rangle \approx O_{\rm mc}+ \frac{\sigma^2_E}{2} \frac{\partial^2 O_{\rm mc}}{\partial E^2},\; \delta O^2\approx \delta O^2_{\rm mc}+(\partial_E O_{\rm mc})^2 \sigma_E^2
 \end{displaymath}
 where $O_{\rm mc}$ is the microcanonical average of $O$ at fixed energy $E$ and $\delta O_{\rm mc}^2$ is the variance of $O$ in the microcanonical ensemble. Thus Eq.~(\ref{widthbeta}) will have implications to fluctuations of a wide class of observables in mesoscopic thermally isolated driven systems.

The authors would like to thank G. Ortiz for the comment related to a cumulant expansion of the Jarzynski equality which plays an important role in the proof. The authors also acknowledge the support of the NSF DMR-0907039 (A.P.), AFOSR FA9550-10-1-0110 (L.D. and A.P.), Sloan Foundation (A.P.), Y.K. thanks the Boston University visitors program for its hospitality.

\section{Methods}\label{extramaterial1}

We now outline a particularly simple derivation of Eqs.~\eqref{widthbeta} and (\ref{real_AB}). A more rigorous derivation based on fluctuation theorems and the unitarity of the evolution is given in the Supplementary Material. Our process consists of many repeated cycles during which the control parameter $\lambda(t)$ is varied in time, returning to its initial value at the end of each cycle. We assume that between the cycles the system reaches a steady state (or a diagonal ensemble \cite{rigol_08} in the quantum language) so that its state is fully characterized by its energy distribution. In ergodic systems this requirement can be satisfied by waiting between cycles a time which is longer than the relaxation time of the system. In non ergodic (integrable) systems this can be achieved by having a long fluctuating time between cycles. This effectively leads to an additional time averaging which is equivalent to the assumption of starting from a diagonal ensemble. (For more details about relaxation to asymptotic states in integrable systems see Ref.~\cite{ap_rmp} and refs. therein). To make this discussion more concrete consider, for example, a compression and expansion of the piston in Fig.~\ref{ovens2}  according to an arbitrary protocol. The gas is allowed to relax between the cycles (when the piston is stationary) at a fixed energy. For a weakly interacting ergodic gas such a relaxation implies that the momentum distribution of individual particles assumes a Maxwell-Boltzmann form together with a randomization of the coordinate distribution. For a noninteracting gas in a chaotic cavity the relaxation implies conservation of the individual energies of each particle and a randomization of the coordinates and directions of their motion. And finally for noninteracting particles in a regular non-chaotic cavity the relaxation implies a randomization of the coordinates within individual periodic trajectories. Therefore, in the beginning of each cycle there are no correlations between positions and velocities of particles within the available phase space.

If we make an additional assumption about ergodicity then the system between the cycles is fully described by the total energy. As we will see later, when we discuss specific examples, this assumption is not always necessary. Assuming that during each cycle a small amount of work is carried out on the system the energy distribution $P(E,t)$ can be described by the Fokker-Planck equation (\ref{FP_main})~\cite{reif}. The easiest way to derive the fluctuation-dissipation relation (\ref{real_AB}) is to note that under very general conditions the only attractor of Hamiltonian dynamics is a flat probability distribution for the occupation of different microstates~(see Ref.~\cite{jarzynski_92} for the classical case and Ref.~\cite{ap_heat} for the quantum case), which is the maximum entropy state. Therefore the energy distribution which is proportional to the many-particle density of states $P_{s}(E)=C\Omega(E)$ should be stationary under the Fokker-Planck equation, implying that the current $J_{s}=-A(E)P_{s}(E)+\frac{1}{2}\partial_{E}(B(E)P_{s}(E))$ is a constant, which vanishes since $P_{s}(E)=0$ for $E$ below the ground-state energy. Finally we use $\beta(E)=\partial_{E}\ln\Omega(E)$ to obtain Eq.~\eqref{real_AB}. For a rigorous derivation of this result and the range of its applicability see Supplemetary Material.  

The relation (\ref{real_AB}) allows us to make general statements about the energy distribution. In particular, the main result of the paper, Eq. (\ref{widthbeta}), immediately follows from Eqs.~\eqref{FP_main} and~\eqref{real_AB} to leading order in an expansion in $1/N$. To see this we first multiply Eq.~(\ref{FP_main}) by $E$ and $E^2$ and integrate over all energies. In this way we obtain the differential equations describing the time evolution of $\langle E \rangle$ and $\sigma^{2}=\langle E^2 \rangle - \langle E \rangle^2$, where angular brackets stand for averaging over $P(E)$:
\begin{eqnarray*}
 &  & \partial_{t}\langle E \rangle=\langle A(E)\rangle\\
 &  & \partial_{t}\sigma^{2}=\left\langle B\right\rangle +2\left(\left\langle A(E) E\right\rangle -\left\langle A(E)\right\rangle \langle E \rangle\right).
\end{eqnarray*}
These two equations can be combined into a single one:
\begin{equation}
\frac{\partial\sigma^{2}}{\partial\langle E \rangle}=\frac{\left\langle B\right\rangle +2\left(\left\langle AE\right\rangle -\left\langle A\right\rangle \langle E \rangle\right)}{\langle A\rangle} \;.
\label{combined}
\end{equation}
If the energy distribution $P(E)$ is narrow, as in the case of large systems, we can evaluate the averages above
using a saddle-point approximation and Eq.~\eqref{real_AB}. Then to order ${\cal O}(N^{-1})$ we find:
\begin{equation}
\frac{\partial\sigma^{2}}{\partial\langle E \rangle} = 2\beta^{-1}(\langle E \rangle)+2\frac{\partial_{E}A(\langle E \rangle)}{A(\langle E \rangle)}\sigma^{2}(\langle E \rangle).
\label{intermediate}
\end{equation}
Integrating this equation immediately yields Eq.~(\ref{widthbeta}).

Let us now comment on the regime of validity of Eq.~(\ref{widthbeta}). The derivation is based on the Fokker-Plank equation, Eq.~\eqref{FP_main}, the generalized fluctuation-dissipation relation, Eq.~\eqref{real_AB}, and the saddle-point expansion in Eq.~\eqref{intermediate}. The validity of the Fokker-Planck equation relies on the assumption that the work distribution is narrow and the average work for cycle is small. More specifically this equation is derived from a cumulant expansion of the Crook's relation up to second order in the work (see Sec.~\ref{extramaterial2} in Supplementary Material for details). Let us here only mention two necessary conditions justifying the Fokker-Planck equation and the fluctuation-dissipation relation Eq.~(\ref{real_AB}): (i) The third (and higher order) cumulant of work per cycle are small $\beta^2(E)\langle w^3\rangle_c\ll \langle w\rangle=A(E)$. (ii) The average work per cycle is smaller than the product of temperature and the specific heat $C_v$: $\beta(E)\langle w\rangle=\beta(E) A(E)\ll C_v$. As explained in the Supplementary Material (sec. \ref{extramaterial2}), if this condition is not satisfied there are corrections of order $\beta A^2(E)/C_v$ to Eq.~\eqref{real_AB}. Finally the saddle-point approximation in Eq.~\eqref{intermediate} is justified if the energy fluctuations in the system are small. This is the case in large or mesoscopic extensive systems.

We also note that our derivation implicitly relies on the assumption of ergodicity within the system. In particular, we are assuming that $P(E)$ is a differentiable function of energy. In integrable systems this is not necessarily the case~\cite{santos_11}. Then the validity of our results should be checked  on a case by case basis. For example, as we showed above, Eq.~\eqref{real_AB} and the corresponding Fokker-Planck equation~\eqref{FP_main} describe the dynamics of a single particle in a cavity. In this case, however, the second term in the RHS of Eq.~\eqref{real_AB} is important and modifies the width of the distribution even if we consider an ensemble of many noninteracting (and therefore non-ergodic) particles.

\clearpage
\newpage
\section*{Supplementary Material}

\setcounter{section}{0}
\section{Derivation of Eq. (4) from the quantum Crook's equality}\label{extramaterial2}

Here we will sketch the derivation of Eq.~\eqref{real_AB} relying only on the unitarity of the dynamics in the quantum case and the incompressibility of Hamiltonian dynamics (i.e. Liouville's theorem) in classical case.  Our proof will be based on the Crooks theorem.  As it was noted in previously (see for example~\cite{Yariv_08_s}), for a closed system obeying classical Hamiltonian dynamics, the Crooks theorem relies on the incompressibility of trajectories in phase space (Liouville's theorem) and microscopic time reversibility. Here we extend this proof to isolated quantum Hamiltonian systems with a discrete spectrum. Our proof of the quantum Crook's relation bears some similarities with that discussed in Ref.~\cite{hanggi_10_s}, and is presented here for completeness. This will emphasize some important properties of the transition matrix, highlight that the Crook's relation does not rely on assumptions related to energy measurements, e.g. at intermediate steps, and extend the fluctuation-dissipation relation \eqref{real_AB} to non-canonical distributions.

Let us assume that a system prepared in a stationary state, described by a diagonal density matrix in the energy basis, undergoes some process described by a unitary operator $U(t)$. According to standard quantum mechanics the density matrix evolves in time according to $\rho(t)=U(t) \rho(0) U^\dagger (t)$. This means that the diagonal elements of the time evolved density matrix in the new energy basis are given by
\begin{equation}
\rho_{nn}(t)=\sum_m U_{nm} \rho_{mm}(0) U^\dagger_{mn}=\sum_m T_{m\to n} \rho_{mm}(0)
\label{master}
\end{equation}
where we used the fact that the initial density matrix is diagonal and introduced the transition probabilities $T_{m\to n}=|U_{mn}|^2$ (see also Ref.~\cite{ap_heat_s}). The matrix $T_{m\to n}$ is doubly stochastic meaning that $\sum_n T_{m\to n}=\sum_m T_{m\to n}=1$. While the first equality is simply the conservation of probability the second is a direct consequence of unitarity. It is easy to see that this equality is violated if there are losses in the system due to e.g. spontaneous emission. Now let us imagine a time-reversed protocol described by the inverse evolution operator $U^{-1}$. From the definition of the transition probabilities it is clear that
\begin{equation}
\tilde T_{n\to m} = T_{m\to n},
\label{crooks_q}
\end{equation}
where $\tilde T_{n\to m}$ refers to the reverse process. Let us comment that the transition probabilities also satisfy detailed-balance, $T_{n\to m}=T_{m\to n}$, in the two following situations: (i) if the Hamiltonian of the system is time-reversal invariant at each moment of time and the protocol is time symmetric so that $U(t)=U(T-t)$, where $T$ is the period of the cycle. (ii) If the transition probabilities during one cycle are small and can be computed within first order in an adiabatic perturbation theory~\cite{ap_heat_s}, i.e. a perturbation theory in a basis evolving with the Hamiltonian (this theory also includes ordinary perturbation theory as a particular limit of small amplitude perturbations). Let us stress that detailed-balance only plays the role in our proof for deriving the subleading $\partial_E B$ correction in the relation~(\ref{real_AB}).

To proceed we use the energy distribution:
\begin{equation}
P(E)=\sum_n \rho_{nn} \delta(E-E_n)
\end{equation}
and relate transition probabilities between energy levels to the transition probabilities between energy shells:
\begin{equation}
T_{E \to E'} = {1\over \Omega(E)}\sum_{n,m} \delta(E-E_n)\delta(E'-E_m) T_{n\to m},
\label{energy}
\end{equation}
where $\Omega(E)=\sum_n \delta(E-E_n)$ is the many-body density of states. The factor $1/\Omega(E)$ ensures conservation of probability: $\int dE' T_{E \to E'}=1$. The master equation (\ref{master}) is then given by
\begin{equation}
P(E')=\int dE\, T_{E\to E'} P_0(E)
\end{equation}

We now multiply both sides of Eq.~\eqref{crooks_q} by $\delta(E-E_n)\delta(E'-E_m)$ and sum over $n,m$ to obtain
\begin{equation*}
\Omega(E) T_{E \to E'}=\Omega(E')\tilde T_{E' \to E}
\end{equation*}
Denoting $E'=E+w$ and using the fact that $\Omega(E)=\exp[S(E)]$ we can rewrite the equation above as
\begin{equation}
T_{E\to E+w}e^{-S(E+w)+S(E)}=\tilde T_{E+w\to E}\;,
\label{crooks-cont}
\end{equation}
which is known as the Crooks relation~\cite{crooks_98_s}.

To prove relation~(\ref{real_AB}) we use Eq.~\eqref{crooks-cont}, and expand the entropy and the transition probability $T_{E\to E+w}$ in $w$:
\begin{equation}
\label{expansion1}
\begin{split}
&S\left(E+w\right)-S\left(E\right)\approx \beta w-\frac{1}{2\sigma_{eq}^{2}}w^{2}\\
&\tilde T_{E+w\to E}=\tilde T_{E\to E-w}+w\partial_{E}\tilde T_{E\to E-w},
\end{split}
\end{equation}
as in the main text we  assume a cyclic process.

Note that when $\langle w \rangle$ is held constant and the system size is increased, the second terms on the RHS of each of the above equations scale as $1/N$. To leading order in $1/N$, integrating Eq.~(\ref{crooks-cont}) over $dw$ we obtain the Jarzynski like relation $\left\langle e^{-\beta w}\right\rangle =1$, where the brackets represent an average over realizations of the process. Note that we did not make any assumption about an initial Gibbs distribution. Taking the logarithm of this Jarzynski relation and performing a cumulant expansion we find
\begin{equation}
\label{lrq}
2\langle w\rangle\approx \beta\langle  w^{2}\rangle_c
\end{equation}
from which Eq.~\eqref{real_AB} is obtained by using
\begin{equation}
\begin{split}
A&=\langle w\rangle, \\
B&=\langle w^2\rangle-\langle w\rangle^2\equiv \langle  w^2 \rangle_c.
\end{split}
\end{equation}
The condition for the validity of this expansion is that the third cumulant of the work is small:
\begin{equation}
\beta^2 \langle w^3\rangle_c\ll \langle w\rangle.
\label{cond1}
\end{equation}

 When the additional assumption of the detailed-balance holds we can use $T=\tilde T$ in Eq.~(\ref{crooks-cont}). We point again that detailed-balance is valid for arbitrary symmetric protocols as well as for non-symmetric protocols, provided that the transition probabilities can be computed within first order of adiabatic perturbation theory.~\footnote{In general situations, which can involve Berry phases, this statement is correct only in the appropriate adiabatic (co-moving) basis.} Since the work per cycle is assumed to be small, it is expected that the transition probabilities are also small and can be computed perturbatively. So the assumption of $T=\tilde T$  between energy shells is likely to be generic. In particular, one can check that it is asymptotically satisfied at high energies for the piston example discussed in the main text even for asymmetric protocols. In this case, integrating Eq.~(\ref{crooks-cont}) using the expansions (\ref{expansion1}) we obtain
\begin{equation}
\left< \exp\left[-\beta w +{w^2\over 2\sigma_{\rm eq}^2}\right]\right>\approx 1-\partial_E\langle w\rangle
\end{equation}
Taking the logarithm of both sides and performing the cumulant expansion of the exponent up to the order $w^2$ we find
\begin{equation}
-\beta \langle w\rangle +{\langle w^2\rangle \over 2\sigma_{\rm eq}^2}+{\beta^2\over 2}\langle \delta w^2\rangle\approx -\partial_E \langle w\rangle.
\end{equation}

It is easy to check that up to order $1/N$ the equations above imply Eq.~(\ref{real_AB}): $2A=\beta B+\partial_E B$ as long as $\langle w\rangle ^2/\sigma_{\rm eq}^2$ is negligible compared to $\beta \langle w\rangle$.
Noting that $\langle w^2\rangle=\langle  w^2\rangle_c+\langle w\rangle^2=B+A^2$ this gives us a necessary condition of validity of relation (\ref{real_AB}):
\begin{equation}
A=\langle w\rangle \ll T C_v \;.
\label{cond2}
\end{equation}
Namely, the work per unit cycle should be small compared to the temperature multiplied by the specific heat.
We note that even though we derived Eq. (\ref{real_AB}) to the order of $1/N$, it is actually correct to all orders in $1/N$. This relation is valid as long as the conditions (\ref{cond1}) and (\ref{cond2}) are satisfied.

 Finally let us discuss extension of the fluctuation-dissipation relation (\ref{real_AB}) to arbitrary distributions. In order to do this we need to weight Eq.~(\ref{real_AB}) with an energy distribution $P(E)$ and integrate over energies. Then it is easy to check that in the Gaussian approximation we find
\begin{equation}
\langle w\rangle\approx {\beta \over 2}\langle w^2\rangle_c+{1\over 2} \left(1-{\sigma^2\over \sigma_{\rm eq}^2}\right) \partial_E \langle w^2\rangle_c.
\label{fluct_diss_rel}
\end{equation}
This relation is clearly a generalization of Eq.~(\ref{real_AB}). In particular, it reduces to Eq.~(\ref{real_AB}) for the microcanonical distribution with $\sigma^2=0$ and it reduces to the conventional result obtained from the cumulant expansion of the Jarzynski relation $\langle w\rangle\approx \beta/2 \langle w^2\rangle_c$ for the canonical distribution $\sigma=\sigma_{\rm eq}$.

\section{Additional Examples}

In this section we will provide two additional examples illustrating validity of our results to a classical and quantum interacting one-dimensional spin chains.

\subsection{An XY model in one dimension}

First we consider an XY-model on a one-dimensional lattice of size $N$. On each lattice site there is a single degree of freedom, which may be viewed as a two-dimensional unit vector. The interaction energy between neighboring sites $i,j$ is $H_{i,j}=1-\cos\left(\theta_{i,j}\right)$, where $\theta_{i,j}=\theta_i-\theta_j$ is the difference between the angles at sites $i$ and $j$. The total Hamiltonian is given by summing over all nearest neighbors the interaction terms $H=\sum_{\left\langle i,j\right\rangle }H_{i,j}$.
To drive the system, we assume that the angle at a specific site $\theta_{i}$ is changed by a small amount $\delta\theta$, which is a fluctuating variable with zero mean, and other sites are unaffected. Such a protocol, for example, describes an interaction of the system with an external fluctuating local magnetic field. The problem can be solved exactly in one-dimension. To do this, note that the change in energy of the system depends only on the two differences $\theta_{i-1,i}$ and $\theta_{i,i+1}$ before the drive. The probability distribution of $\theta_{i,i+1}$ to order ${\cal O}(1/N)$ is given by
\begin{equation}
\rho(\theta_{i,i+1})\propto\exp[\beta(E)\cos(\theta_{i,i+1})].
\end{equation}
Using this expression, it is straightforward to calculate
the values of the average work and its fluctuations $A(E)=\left\langle w\right\rangle , B(E)=\left\langle w^{2}\right\rangle_c$ to order $\delta\theta^{2}$:
\begin{equation}
A=\left\langle \delta\theta^{2}\right\rangle \left(1-\frac{E}{N}\right),\; B=\frac{2A}{\beta}.
\label{AB_xy}
\end{equation}
Here to relate $E$ and $\beta$ we used the expression for the energy \cite{Pathria_s} $E/N=1-\left[I_{1}\left(\beta\right)/I_{0}\left(\beta\right)\right]$, where $I_{n}$ denotes a modified Bessel function of the first kind of order $n$.

\begin{figure}
\includegraphics[width=0.9\columnwidth]{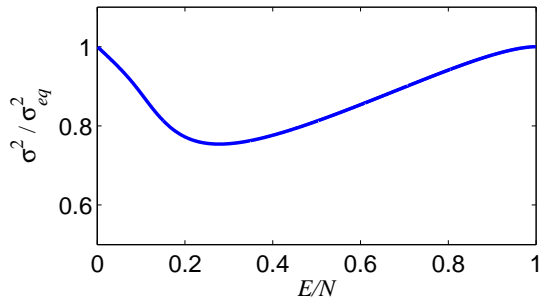}
\caption{The ratio $\sigma^2(E)/\sigma_{eq}^2(E)$ as a function of $E/N$ for a one-dimensional XY-model. The initial conditions are $E_0=0$ and $\sigma^2_0=0$.}
\label{fig:XY}
\end{figure}

Substituting relations (\ref{AB_xy}) into Eq.~(\ref{widthbeta}) and numerically integrating we obtain $\sigma^2(E)/\sigma_{eq}^2(E)$ as a function of $\beta(E)$. The results are shown in Fig.~\ref{fig:XY}. Note that there are two regimes. In the low energy regime $A$ is to lowest order constant, $s=0$, and $\alpha=1$, which gives $\eta=-2$, and $\sigma^2/\sigma_{eq}^2 \simeq 1$, see Table~\ref{summary}. The next order correction for small $E/N$ can be obtained using Eq.~(\ref{AB_xy}) and gives: $\sigma^2/\sigma_{eq}^2 \simeq 1 - E/N$. Then at high energies, close to infinite temperature, $E/N=1$ and again $\sigma^2/\sigma_{eq}^2 \to 1$. This is due to the finite phase-space available to this system, which allows the system to reach a stationary distribution with $\beta=0$ at a finite total energy density ($E/N=1$).

\subsection{The transverse-field Ising model in one dimension}

Finally we consider a quantum transverse-field Ising model in one dimension described by the Hamiltonian:
\begin{equation}
\mathcal{H}=-\sum_j [g \sigma^x_j+\sigma^z_j\sigma^z_{j+1}]+H',
\label{ising_model}
\end{equation}
 where $H'$ is a weak perturbation breaks integrability of the system but does not affect the dynamics during the driving protocol. For example, this perturbation can be a weak second nearest-neighbor spin-spin interaction. The assumption of smallness of $H'$ is only important to make explicit analytic calculations. For simplicity we will consider the domain of non-negative values of the transverse field $g$. This system undergoes a quantum phase transition at $g=1$ \cite{Sachdev_s}. Using a Jordan-Wigner transformation the Hamiltonian assumes a quadratic form in the fermionic operators and can be fully diagonalized performing a Bogoliubov rotation in momentum space~\cite{Sachdev_s}. The final Hamiltonian reads:
\begin{equation}
\mathcal{H}=\sum_k \epsilon^g_k (\gamma_{k}^{\dagger}\gamma_k-\frac{1}{2}),
\end{equation}
where
\begin{displaymath}
\epsilon^g_k=2 \sqrt{1+g^2-2g\cos(k)}\approx 2\sqrt{(1-g)^2+k^2}
\end{displaymath}
and $\gamma_{k}^{\dagger},\,\gamma_{k}$ are quasi-particle creation and annihilation operators. Here in order to simplify analytic expressions we linearized the spectrum by taking $g-1\ll 1$ and the relevant momenta are much smaller than the ultra-violet cutoff given by the lattice: $|k|\ll \pi$. The ground state of this Hamiltonian, which is annihilated by all quasi-particle operators $\gamma_k$, is factorized into momentum sectors. The excited states can be obtained by applying various combination of operators $\gamma_k^{\dagger}$ to the ground state. If the external time-dependent perturbation is spatially uniform then due to momentum conservation only the excited states corresponding to pairs of excited quasi-particles with opposite momenta obtained by applying $\gamma_k^{\dagger}\gamma_{-k}^{\dagger}$ to the ground state participate in the dynamics. Moreover excitations to different momentum states are independent and the problem effectively splits into a collection of two levels-systems. Let us emphasize that this system does not become classical even in the infinite temperature limit. In disordered systems this feature was explored e.g. in the context of the many-body localization in Ref.~\cite{oganesyan_07}.

We now perturb the system by changing the amplitude of the transverse-field $g(t)=g_1+\delta\, t (1-t/\tau)$, $t\in[0,\tau]$ from its initial value, $g_1$, to an intermediate value, $g_2=g_1+\delta \tau/4$ and then back to $g_1$. Here the parameter $\delta$ sets the velocity of the quench and $\delta \tau/4$ sets its amplitude. As a result of this process the occupation of the energy levels will change. Because different momentum modes are effectively independent from each other we will consider each two-level system separately. The presence of the weak integrability breaking perturbation ensures a Fermi-Dirac redistribution of the energy among different modes between different cycles. It is similar to the effect of a weak interaction between particles leading to a Maxwell-Boltzmann single-particle distribution in the piston example or an assumption of weak coupling to the rest of the system in the single oscillator example discussed in the main text. The ergodicity of small weakly nonintegrable one-dimensional systems was recently tested numerically to a very good accuracy for hard-core bosons and fermions, which are closely related to the transverse field Ising model (see e.g. Ref.~\cite{santos_10}).

Under these assumptions each cycle starts from the Fermi-Dirac distribution in each momentum mode. To avoid extra complications related to the singularities of the transition probabilities at the critical point we will additionally assume that dynamics occurs only in one phase, say $g>1$ (at finite temperatures we are interested in, this assumption can be further relaxed). For slow quenches, if the rate $\delta$ is small: $\delta\ll (g_1-1)^2$, such that the adiabaticity condition $\delta\ll \epsilon_k^2$ is satisfied for all modes, the dynamics can be solved analytically. In particular, in Ref.~\cite{degrandi_09} it was shown, that under these conditions the transition probability between the ground and excited states is approximately equal to (see Eqs.~(20) and (87) in Ref.~\cite{degrandi_09})
\begin{equation}
\label{T_k}
p_k\approx {1\over 32} {\delta^2 k^2\over (k^2+(g_1-1)^2)^3}=\frac{2\delta^2 k^2}{(\epsilon_k^{g_1})^6}.
\end{equation}
Note that there is an additional contribution to Eq.~(\ref{T_k}) which is rapidly oscillating at a frequency $\omega\sim (g_2-g_1) \tau$ (see Eq.~(20) in Ref.~\cite{degrandi_09}) and averages to zero either because of adding contributions from different momentum modes or because of slight fluctuations of the quench time $\tau$ from cycle to cycle. Actually the addition of the oscillating term to Eq.~(\ref{T_k}) will at most double the transition probability. As we will see below our results are not affected by the precise form of $p_k$ as long as $p_k$ is small, which is controlled by $\delta$ and decays sufficiently fast with momentum $k$ so that the work distribution remains sufficiently narrow.

\begin{figure}
\includegraphics[width=0.9\columnwidth]{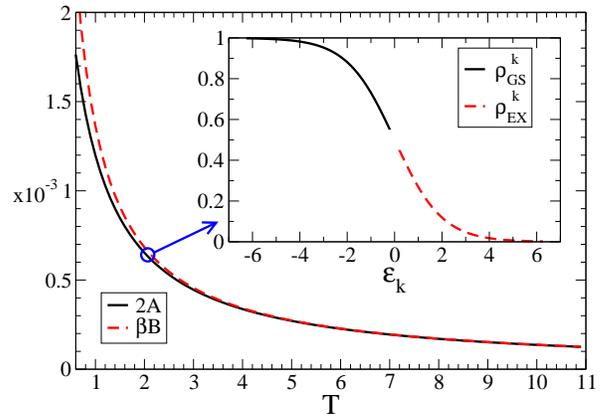}
\caption{The validity of the Einstein-like relation for the transverse-field Ising model as a function of the initial temperature.
The transition probability are calculated as in Eq.~\eqref{T_k} with $\delta=0.05$ and $g_1=1.1$.
In the inset we show the initial occupancy of energy level at $T=2.0$.}
\label{Isingplot}
\end{figure}

With these transition rates, the master equation for the occupation probabilities of different momentum states becomes particularly simple:
\begin{equation*}
\left(\begin{array}{c}
\tilde{\rho_k}^k_{Gr}\\
\tilde{\rho_k}^k_{Ex}\end{array}\right)=
\left(\begin{array}{cc}
1-p_k & p_k\\
p_k & 1-p_k\end{array}\right)
\left(\begin{array}{c}
\rho^k_{Gr}\\
\rho^k_{Ex}\end{array}\right),
\end{equation*}
where $\rho^k_{Ex}$ and $\tilde{\rho}^k_{Ex}$ refer to the probabilities of having a pair of fermions with momenta $k$ and $-k$ before and after the quench respectively and likewise, $\rho^k_{Gr}$ and $\tilde{\rho}^k_{Gr}$ are the probabilities to have no fermions in the $k$ and $-k$ momentum mode. Due to rethermalization of the system between driving protocols we have
\begin{equation}
\rho^k_{Gr}={\exp[\beta \epsilon_k^{g_1}]\over 2\cosh[\beta\epsilon_k^{g_1}]},
\rho^k_{Ex}={\exp[-\beta \epsilon_k^{g_1}]\over 2\cosh[\beta\epsilon_k^{g_1}]}
\end{equation}
Actually these expressions are somewhat modified due to presence of the inert modes with one fermion in either $k$ or $-k$ mode, but this modification does not affect our conclusions since the inert modes do not participate in transitions only effectively reducing the number of active modes by a factor of two in the regime of interest. From the master equation and the probabilities of occupying initial states above we find the average work and its second moment for individual momentum modes:
\begin{equation}
\label{W-W2}
\begin{split}
\langle w_k \rangle&=2\epsilon^{g_1}_k p_k \tanh(\beta \epsilon^{g_1}_k)\\
\langle w^2_k \rangle&=(2\epsilon^{g_1}_k)^2 p_k
\end{split}
\end{equation}
From this equation we deduce the relation (\ref{real_AB}) between $A_k=\langle w_k\rangle$ and $B_k=\langle w_k^2\rangle_c$: $2A_k\approx \beta B_k$ is satisfied provided that $p_k\ll 1$, which is the case for sufficiently slow quenches, and $\beta \epsilon_k^{g_1}\ll 1$ which means that the temperature is big compared to the initial energy of the fermions. The latter condition is always satisfied if the temperature is large compared to the gap in the system and that the relevant excited modes correspond to small momenta, which is the case for sufficiently slow quenches. If these conditions are fulfilled then Eq.~(\ref{real_AB}) is satisfied for the total work and its variance:
\begin{equation*}
A=\sum_{k>0} A_k,\quad B=\sum_{k>0} B_k.
\end{equation*}
In Fig.~\ref{Isingplot} we show the calculated values of $2A$ and $\beta B$ for different initial temperatures
for the protocol with $g_1=1.1$ and $\delta=0.05$ (the time $\tau$ drops out from the answer if $\delta \tau^2\gg g_1$).
As the temperature increases the work per cycle, $A$, decreases due to fermion anti-bunching and the relation (\ref{real_AB}) is satisfied to a very good accuracy as soon as temperature becomes much bigger than the gap.

Let us make now a number of remarks on validity of our results and the importance of the protocol in quantum systems. The general analysis and derivation of the Fokker-Planck equation in the main paper as well as the relation (\ref{real_AB}) in Sec.~\ref{extramaterial2} of this Supplementary Material (see Eq.(\ref{lrq})) are based on a quantum formalism. Therefore as soon as the conditions $\beta \langle w\rangle\ll C_v$ and $\beta^2\langle w^3\rangle \ll \langle w\rangle$ are satisfied Eq.~(\ref{real_AB}) should work. And indeed we illustrated this here for a particular example of a transverse field Ising model. Instead of this model we could get e.g. fermions with two-body interactions and would come to similar conclusions. The smallness of work can be controlled by doing sufficiently slow quenches or by quenches of small amplitude similar to the classical case. There is an important subtlety related to the second condition of the smallness of the third and higher cumulants of work which distinguishes quantum and classical systems. In particular, in classical situations it is impossible to give a large energy to a system during a cycle unless the external parameter changes suddenly. E.g. in a deforming cavity example during each collision with a wall the particle can gain at most velocity $2V$, where $V$ is the velocity of the wall. This means that in classical situations the work distribution is typically bounded and the smallness of average work usually implies smallness of its cumulants. In quantum systems the situation is very different. Namely in any protocol it is possible to give the system arbitrarily large energy. For smooth protocols, where the external parameter changes analytically in time the transition probability to high energy states decreases exponentially like in the conventional Landau-Zener problem and such transitions do not affect cumulants of work. However, for non-analytic protocols where e.g. amplitude, velocity (like in our case) or acceleration experience a discontinuity the transition probability to high energy states decreases only algebraically with energy like in Eq.~(\ref{T_k}) (see Ref.~\cite{degrandi_09} for more details). This means that high enough cumulants of work necessarily diverge. Of course real experimental protocols are always smooth and these divergencies are cutoff, however the degree of smoothness introduces a new quantum scale into the dynamics. Thus the work distribution can be effectively wide or narrow in the sense of satisfying the condition $\beta^2\langle w^3\rangle \ll \langle w\rangle$ depending on the ratio of this new energy scale and temperature.

Since this discussion is not directly related to this paper we postpone it until future work. Here we only explicitly analyze another protocol where the coupling changes quadratically in time: $g(t)=g_1+(\varepsilon/2)\, t^2 (1-t/\tau)^2$. In this case in the slow limit $\varepsilon\lesssim (g_1-1)^3$ instead of Eq.~(\ref{T_k}) we get (see Ref.~\cite{degrandi_09} for details):
\begin{equation}
p_k\approx {1\over 2} {\varepsilon^2 k^2\over (\epsilon_k^{g_1})^8}.
\label{T_k1}
\end{equation}

\begin{figure}[ht]
\includegraphics[width=0.9\columnwidth]{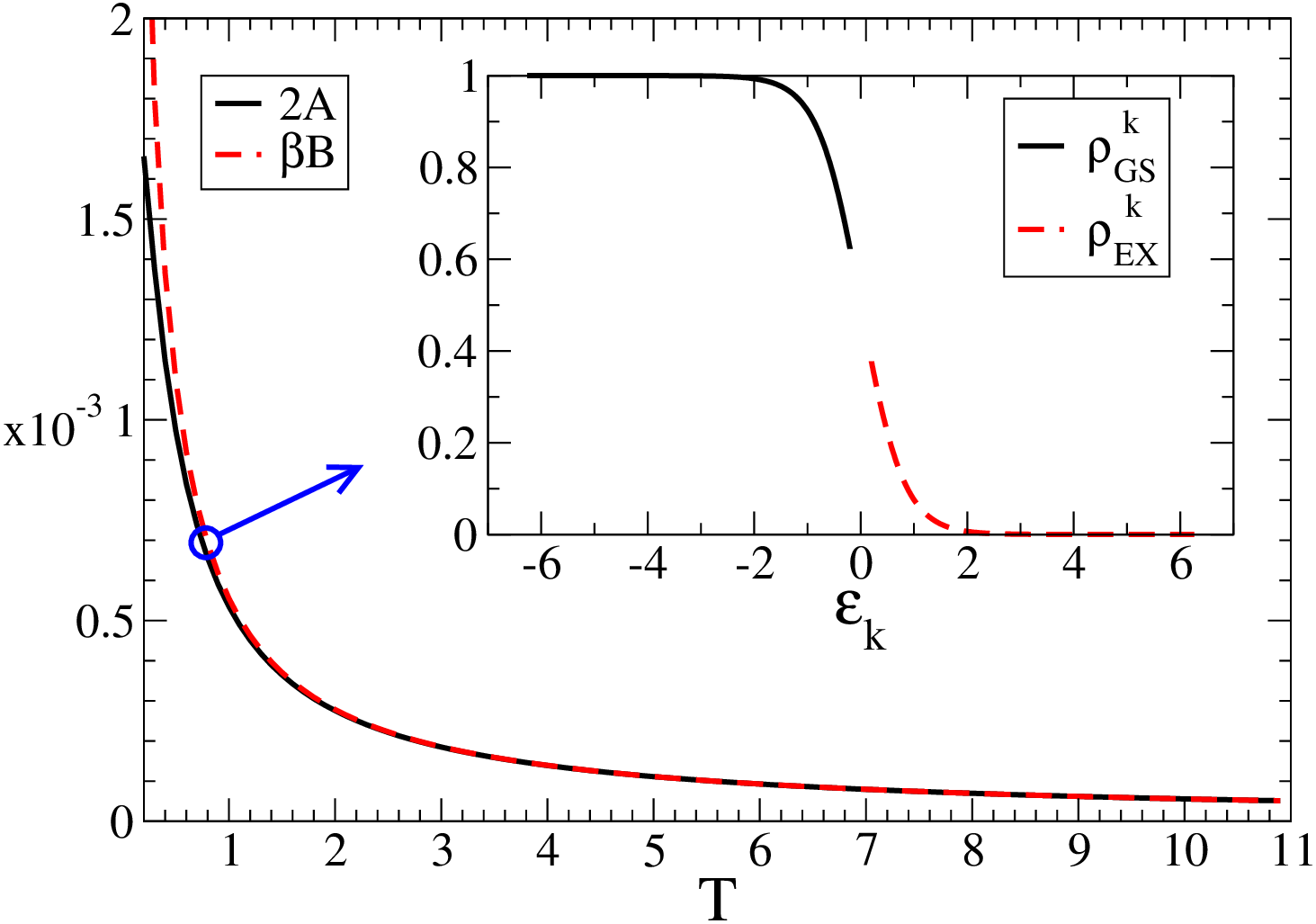}
\caption{Einstein-like relation for the transverse-field Ising model as a function of the initial temperature.
The transition probability are calculated as in Eq.~\eqref{T_k1} with $\epsilon=0.024$ and $g_1=1.1$.
In the inset we show the initial occupancy of energy level at $T=0.8$.}
\label{Isingplot1}
\end{figure}

As we see indeed the transitions to the higher energy states are suppressed even more than for the linear protocol so that more cumulants of work now converge at small energies. As a result the relation (\ref{real_AB}) is satisfied even at smaller temperatures, see Fig.~\ref{Isingplot1}.
In Fig.~\ref{Isingplot1} we show the values of $2A$ and $\beta B$ obtained using Eq.~\eqref{T_k1} with $g_1=1.1$ and $\epsilon=0.024$ (the time $\tau$ drops out from the answer if $\varepsilon \tau^4\gg g_1$). With this choice of the parameters the peak value of $p_k$ in the two protocols
(linear, Eq.~\eqref{T_k}, and quadratic, Eq.~\eqref{T_k1}) are equal. The Einstein-relation, $2A=\beta B$, is satisfied when the relative error $(B/T-2A)/2A$ becomes less than $5\%$. With this definition the Einstein-relation is satisfied at $T>2$ for the linear and $T>0.8$ for the quadratic protocol.

In the regime of validity of the fluctuation-dissipation relation~(\ref{real_AB}) we can expand expression \eqref{W-W2} to first order in $\beta$ to obtain $A\sim \beta\sim T^{-1}$ for both the linear and quadratic protocols described above. Moreover, if the temperature $T$ is in turn much smaller than the cutoff energy scale given by $J$ (which is nothing but the Fermi energy in the fermion representation) then $E\sim T^2$, which is the case for weakly interacting fermions if $T\ll E_F$. From these considerations it immediately follows that $\alpha=1/2$, $s=-1/2$ and $\eta=-5/2$. Therefore $\sigma^2/\sigma^2_{eq} \rightarrow 2/5$. At high energies, i.e. close to the infinite temperature limit, $A\to 0$ and $\sigma^2/\sigma^2_{eq}$ must tend to $1$. As in the XY-model this is due to the finite size of the Hilbert space (see discussion after Eq.~\eqref{AB_xy}). This example shows that for weakly interacting fermions (to which the Ising spin chain is equivalent) it is easy to get distributions significantly narrower than canonical without any special fine tuning. We expect this to be generic result following from the Pauli exclusion principle and suppression of transitions at higher temperatures (see Ref.~\cite{degrandi_09} for additional discussion).

\end{document}